\documentclass[twocolumn]{aastex631}

\usepackage{CJK}

\usepackage{relsize}
\definecolor{red2}{rgb}{0.89, 0.02, 0.17}

\usepackage[autostyle]{csquotes}
\usepackage{subfiles}
\usepackage{soul}
\usepackage{float}
\graphicspath{{./}{Plots}}
\usepackage{amsmath}
\usepackage{amsmath}	
\usepackage{amssymb} 
\usepackage{multirow}
\usepackage{booktabs}
\usepackage{hyperref}
\usepackage{xurl}

\usepackage{graphics}
\newcommand{\msun}{\text{M}_\odot} 
\newcommand{\lsun}{\text{L}_\odot} 

\shortauthors{Cruz-Cruz et al.}

\begin{document}

\title{The Progenitor of the S147 Supernova Remnant}



\shortauthors{Cruz-Cruz et al.}
\correspondingauthor{Elvira Cruz-Cruz}
\email{cruz-cruz.1@buckeyemail.osu.edu}
\author[0000-0002-7260-8774]{Elvira Cruz-Cruz}
\affil{Department of Astronomy, The Ohio State University, 140 W 18th Ave, Columbus, OH 43210 USA}

\author[0000-0001-6017-2961]{Christopher S. Kochanek}
\affil{Department of Astronomy, The Ohio State University, 140 W 18th Ave, Columbus, OH 43210 USA}
\affiliation{Center for Cosmology and AstroParticle Physics (CCAPP), \\ The Ohio State University, 191 W. Woodruff Avenue, Columbus, OH 43210, USA}



\begin{abstract}
The supernova remnant (SNR) S147 contains the pulsar PSR J0538+2817 and a likely unbound binary companion, HD 37424. It is the only good Galactic candidate for a binary unbound by a core-collapse supernova (SN). Using Gaia DR3 parallaxes and photometry, we select the stars local to SNR S147 in a cylinder with a projected radius of $100$ pc and a parallax range of $0.614 < \varpi < 0.787$ mas (a length of $\simeq 360$ pc). We individually model the most luminous of these stars. The two most luminous single stars are the unbound binary companion, HD 37424, and HD 37367, with estimated masses of $(13.51\pm0.05) \msun$ and $(14.30\pm0.09) \msun$, respectively. The two most luminous binary systems are the spectroscopic binary HD 37366 and the eclipsing binary ET Tau that have primary masses of $(20.9\pm0.12) \msun$ and $(16.7\pm0.09) \msun$, respectively. We model the Gaia color-magnitude diagram (CMD) of this local stellar population using both single stars and a model consisting of noninteracting binaries using Solar metallicity \texttt{PARSEC} isochrones. For both models, the estimated age distributions of the $439$ $M_{G} < 0$ mag stars favor a high mass progenitor of $21.5\msun-41.1\msun$ for the SN.
\end{abstract}


\section{Introduction} \label{sec:intro} 

Understanding the progenitors and properties of supernova remnants (SNR) is essential for understanding massive star evolution, supernova mechanisms, and the formation of compact objects. Most stars more massive than $\sim 8\msun$, have short lifetimes, evolve, and end their lives in core-collapse supernova (ccSN) explosions, leaving behind SNRs and a compact object, either a neutron star or a black hole. Some of the most well-studied supernova remnants contain pulsars, including the Crab, Vela, and S147 (PSR J0538+2817). None of these systems are presently binaries, but most massive stars, about  $70\%$ in \cite{Sana2012}, are initially in binaries or higher-order systems \citep{Sana2012,Moe&DiStegano2013closebinarymassivestars}. We know of three compact object binaries in Galactic SNRs, two neutron star wind accreters \citep[$1$FGL J$1018.6 - 5856$, HESS J$0632 + 057$;][respectively]{Corbetbinary2011,Hinton-Hess2009ApJ} and the black hole Roche-lobe overflow system \citep[SS 433;][]{SS433-distance-2004, SS4332004,SS433Russia2025}.

Massive binary star systems can interact as they evolve, modifying the nature of the resulting supernova. The explosions usually unbind the binary \citep[e.g.,][]{Blaauw1961, TaurisTakens1998, BPSBelczyBulik1999,BPSBelczynskiStarTrack2008,Eldridge2011BPS,Renzo2019}. It was thought that these unbound binary companions could be found as OB stars with high space velocities \citep[OB runaway stars,][]{Blaauw1961}, but recent binary population synthesis studies find that the typical velocities are unremarkable \citep{Renzo2019,Wagg2025}.

Galactic SNRs can be used to constrain the statistics of bound and unbound binaries \citep{Dincel2024binarycompanions, Kochanek2023SearchforUnboundStellarCompanions, Kochanek2019stellarbinaries, Kochanek2018notstellarbinariesatdeath, Boubert2018GaiaRunawaycompanions, Boubert2017binarycomapnionsGaia, FraserBourbert2019survivingbinarycompanions, Lux2021searchrunawaystarsGalacticSNRs} and triples \citep{Kochanek2021SEDfits, BarbozaKochanek2024}. While the uncertainties are still large, \cite{Kochanek2023SearchforUnboundStellarCompanions} found that $74\%$ ($55\%-87\%$) are not binaries at death, $13\%$ ($5\%-26\%$) are in bound binaries, and $12\%$ ($3\%-29\%$) are in unbound binaries. Interestingly, all three of the known binaries are interacting binaries, with a $<8\%$ limit on SNRs having a surviving, but non-interacting binary \citep{Kochanek2023SearchforUnboundStellarCompanions}.

One well-developed method for estimating the progenitor masses of SNRs is studying the stellar population near the supernovae and SNRs in external galaxies. The supernova progenitor mass is estimated from the local star formation history found by modeling the color-magnitude diagram (CMD) of the surrounding stars. This technique has been applied to ccSN remnants in galaxies such as M31, M33, and the Small Magellanic Cloud \citep[][]{Jenningsprogmass2012, Auchettl2019}. \cite{Murphyprogmass2011} also did this with SN 2011dh   and \cite{Williamsprogmass2014} and \cite{Diazprogmassdistr2021} have done this to calculate the progenitor mass distribution of 17 (22) historic ccSNe. 

Using this method in the Galaxy only became possible with Gaia \citep{GaiaCollab2016, GaiaCollab2023j}. \cite{Kochanek2022}, \cite{MurphyNewVelaProg2025}, and \cite{cruzcruz2024} have successfully applied it to the surrounding stellar populations of the Vela pulsar and the Crab pulsar, respectively. \cite{Kochanek2022} estimated Vela's progenitor mass of $\leq 15 \msun$. \cite{MurphyNewVelaProg2025} also favor lower masses and argue that the local stellar population likely requires binary interactions to explain its properties. \cite{cruzcruz2024} found that the progenitor of the Crab was likely to have been a low-mass star, comparable to an extreme AGB star or a binary merger product of lower-mass stars.

Here we consider the SNR S147, which has the pulsar PSR J0538+2817 located near its center \citep{SunS147Pulsar1995, DenoyermaposradiospecS1471974, Ng2002PulsarToriSpinKick}. The parallax and transverse velocity of PSR J0538+2817 are $1.47^{+0.42}_{-0.27}$ kpc and $V_{\bot} = 400^{+144}_{-73}$ km $s^{-1}$ \citep{Ng2007OriginMotionS147}. \cite{Ng2007OriginMotionS147} argue that the pulsar progenitor was likely a runaway star from a nearby cluster, potentially within NGC 1960 (M36). \cite{Dincel2015DiscoveryOBRunawayS147} analyzed the kinematics of stars surrounding SNR S147 and found that HD 37424 is moving away from the center of the SNR with a velocity of $74\pm8$ km$s^{-1}$ and that its past trajectory would intersect that of the pulsar on a time scale consistent with the age of the SNR. \cite{Dincel2015DiscoveryOBRunawayS147} proposed HD 37424 as the unbound binary companion of PSR J0538+2817's progenitor. Due to the absence of O stars near the remnant, they argued for an upper mass limit of $20-35$$\msun$ for the supernova progenitor.

A major challenge to applying the star formation method to Galactic SNRs is that they do not have good distance measurements. Recently, \cite{Kochanek2024DistanceS147} successfully estimated the distance to the SNR S147 by looking for high-velocity CaII or Na I absorption lines in stars projected on the SNR as a function of distance. They used the appearance of the high velocity lines to measure a distance of $1.37_{-0.07}^{+0.10}$ kpc to the remnant, which is consistent with the distance to the pulsar of $1.46_{-0.34}^{+0.64}$ kpc \citep[parallax of $0.72 \pm 0.12$mas,][]{S147distance-Ng2007, Chatterjeepulsarparallax2009} and HD 37424's photogeometric distance of $1.449$ kpc ($1.399–1.501$ kpc at $1 \sigma $) from \cite{Bailer-JonesGaiaEDR32021}. These well-measured stellar distances to S147 allow us to apply the stellar population analysis method with Gaia DR3.

In Section \ref{sec:stellarpop} we describe the selection of the stars surrounding PSR J0538+2817 and HD 37424, and the spectral energy distributions (SEDs) of the most luminous stars. In Section \ref{sec:analysis} we introduce the single and non-interacting binary stellar density models created to analyze the star formation history of S147's local stellar population, and then explain how we estimate the likely mass of the progenitor star. In Section \ref{sec:results} we discuss our results and in Section \ref{sec:discussion} we summarize our findings and discuss the implications of our results.

\section{Stellar Population Selection} \label{sec:stellarpop}

We select stars using Gaia DR3 \citep{GaiaCollab2016, GaiaCollab2023j} and Astroquery \citep{astroquery2019}. Each star had to have a parallax and all three Gaia magnitudes (G, $R_{P}$ and $B_{P}$). We use S147's position, RA: $84.75^\circ$ and DEC: $27.83^\circ$, from \cite{Green2009} as the center. We initially selected stars within $\theta = \sin^{-1}(R/D) = 5.24^{\circ}$, where $D = 1370$ pc and $R = 125$ pc, and parallaxes $0.5 < \varpi < 1$ mas. We use an apparent magnitude limit of $G < 15$ to include stars with absolute magnitudes $M_{G}\leq 4.3$ mag and masses that are $M\gtrsim1 M_{\odot}$.

\begin{figure}
    \centering
    \includegraphics[width=1\linewidth]{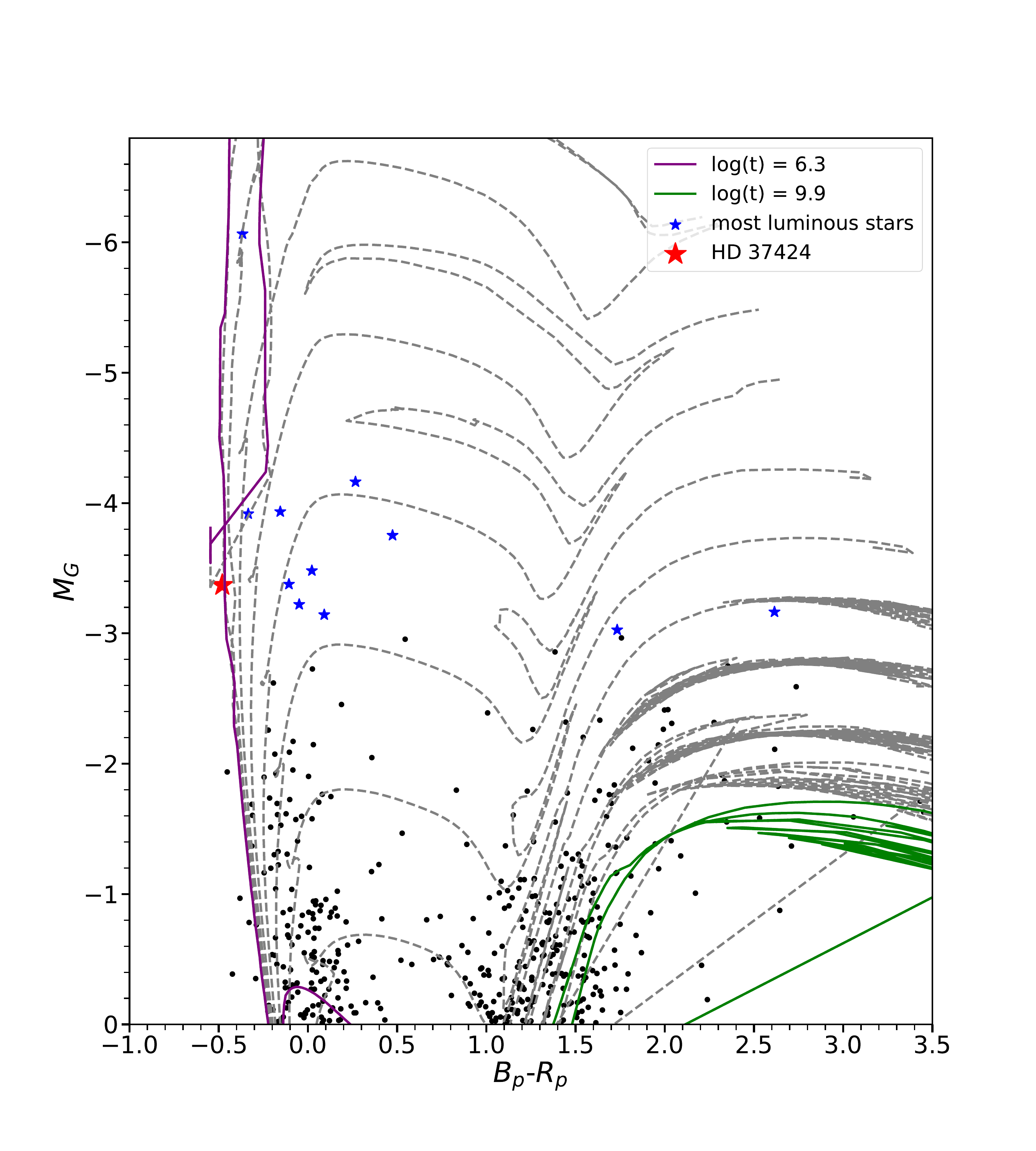}
    \caption{Extinction corrected Gaia DR3 color-magnitude diagram of the stars local to S147 SNR. HD 37424 is the large red star and the other 12 ($M_{G} < -3$ mag) stars are shown with blue small stars. Solar metallicity \texttt{PARSEC} isochrones are shown from $\log_{10}(t/yr) = 6.3$ (solid purple, top left) to $\log_{10}(t/yr) = 9.9$ (solid green, lower right) in steps of 0.3 dex.}
    \label{fig:S147_CMD} 
\end{figure}

We prune this truncated cone to a truncated cylinder with a projected radius of $R = 100$ pc from S147. The Gaia parallax for HD 37424 \citep{Bailer-JonesGaiaEDR32021} lies modestly in front of the S147 distance from \cite{Kochanek2024DistanceS147}.
We chose a minimum parallax corresponding to being $100$pc in front of HD 37424 and a maximum parallax corresponding to $100$pc behind the S147 distance. This leads to a sample of $ 8583$ stars with $d_{\perp} < R$ and parallaxes between $0.614 < \varpi < 0.787$ mas \citep{GaiaCollab2023j}. The parallax range corresponds to a distance range of $\simeq 360$pc. For each star, we used the estimated extinction from the 3-dimensional (3D) \texttt{combined19 mwdust} models \citep{Bovy2016} that are based on \cite{Green2019} near S147. Combining this with the \cite{Bailer-JonesGaiaEDR32021} and distances we obtain the extinction corrected colors ($B_{P}-R_{P}$) and absolute magnitudes ($M_G$). 
We find five stars that are very red, $B_{P}-R_{P} > 3.5$, and place them on the right edge of the observed CMD. The final sample consists of $439$ stars with $-8< M_{G} < 0$ and $-1.0 < B_{P}-R_{P}<3.5$ mag. There are no stars that have $M_{G} < -8$ mag. The CMD of the selected stars is shown in Figure \ref{fig:S147_CMD}. 
For comparison, we show Solar metallicity \texttt{PARSEC}
\footnote{See \url{http://stev.oapd.inaf.it/cmd} for public access to \\ \texttt{PARSEC} stellar tracks and isochrones.}
isochrones in age steps of 0.3 dex that span $10^{6.3}$ to $10^{9.9}$ years \citep[e.g.,][]{PARSEC2012Bressan, PARSEC2013MarigoAGB, PARSEC2020PastorelliTPAGB, Chen2014parsec, Chen2015Parsec, Tang2014parsec, Marigo2017parsec, Pastorelli2019parsec}.

\begin{table*}
    \centering
    \caption{SED fit results of the most luminous stars local to S147, those with a $^\star$ are from \cite{Kochanek2021SEDfits}.}
    \renewcommand{\arraystretch}{1.5}
        \begin{tabular}{c c c c c c c c c}
        \toprule
        Star& $\chi^{2} / N_{dof}$  & $\log(T_{\ast}) [K]$ &$\log(L_{\ast}) [\lsun]$ & $M_{\ast} [\msun]$  & $\log(t) [yr]$ & Sep [pc] & Comments \\ [0.5ex]
        \toprule
         HD 37367 & $0.69$ & $ 4.18\pm 0.04$ & $4.69\pm0.137$ & $13.53-15.15$ & $7.10-7.17 $ & $ 31.762 $ & B2 IV-V \\ [0.5ex]
         BD +26935 & $ 0.90$ & $3.59\pm0.01$ & $3.30\pm0.01$& $3.33-3.53$ & $8.46-8.53$ & $43.119$ & M2/3, Spectroscopic Binary \\ [0.5ex]  
         HD 37366 &  $5.74$ & $4.53\pm 0.01$ & $4.81\pm0.03$ & $ 20.0-21.7$ & $ 6.40-6.67$ & $68.273$ & O 9.5 IV, Spectroscopic Binary \\ [0.5ex]
         HD 38749$^\star$ & $1.05$ & $3.91\pm0.02$ & $3.65\pm0.073$ & $6.54-7.29$ & $7.65-7.75$ & $81.535$ &  A5 \\  [0.5ex]
         HD 36665$^\star$ &  $1.24$  & $4.33\pm0.05$  & $4.41\pm0.155$ & $10.04-13.72$ & $7.13-7.38$ & $24.485$ &  B1, Be star\\  [0.5ex]
         HD 38017A$^\star$ & $1.02$  & $4.24\pm0.03$ & $4.24\pm0.065$ & $9.67-10.59$ & $7.34-7.41$ & $87.163$ &  B3V, Visual Binary\\   [0.5ex]
         HD 38658$^\star$ &  $0.83$ &  $4.22\pm0.03$  & $4.57\pm0.109$ & $11.99-14.17$ &  $7.14-7.25$  & $55.248$  & B3II \\  [0.5ex]
         V399 Aur$^\star$ &  $2.39$ &  $3.53\pm0.00 $& $2.86\pm0.016$  & $0.99 - 1.40$ & $9.54-10.07$ & $83.987$  & M2/3, LPV \\ [0.5ex]
         HD 246370$^\star$ & $0.23$ & $3.85\pm0.03$  & $3.38\pm0.081$  & $5.60-6.26$ & $7.79-7.90$ & $48.084$ &  G5 \\ [0.5ex]
         ET Tau & $1.68$ & $4.46\pm0.04$ & $4.68\pm0.09$  & $15.12-18.19$ & $6.79-7.03$ & $16.804$ & B8, Eclipsing Binary \\ [0.5ex]
         HD 37424$^\star$ & $1.08$ & $4.45\pm0.04$  & $4.28\pm0.121$  &  $11.75-15.48$ & $6.30-7.10$ &  $4.557$ &  B9 \\ [0.5ex]
         HD 248666 & $0.65$ & $4.33\pm0.02$ & $4.19\pm0.07$ & $9.19-11.09$ & $7.27-7.45$ & $91.298$ & B\\[0.5ex]
        \hline
    
        \end{tabular}
    \label{tab:sedlumstars}
\bigskip
\end{table*}

We fit the SEDs of the 12 luminous stars with $M_{G} < -3$ mag in our sample. Among these there are four binary systems, two of which are spectroscopic binaries, BD +26935 with a spectral type of M$2/3$ and HD 37366 with an O-type primary, one visual binary (HD 38017A, B3V) and one eclipsing binary (ET Tau, B8). \cite{Kochanek2021SEDfits} had previously fit the spectral energy distributions of 7 of these stars to estimate luminosities, temperatures, and extinctions. Here we model the SEDs of the remaining stars, HD 37367, BD +26935, HD 248666, and HD 37366, and we redo the SED fit for ET Tau. As in our previous work \citep{cruzcruz2024} we use \texttt{DUSTY} \citep{ElitzuretalDUSTY2001} inside a Markov Chain Monte Carlo (MCMC) driver to optimize SED fits and their uncertainties following methods of \cite{Adamsetal2017a} and \cite{Kochanek2022}. We use \texttt{MARCS} \citep{MARCSGustafsson2008} or \cite{CastelliandKurucz2003} model atmospheres. We searched Vizier \citep{vizier2000} to obtain fluxes for the stars. We use UV fluxes from \cite{ThompsonetalUV1978} or \cite{WesseliusetalUV1982} if available, and optical magnitudes from \cite{Johnsonetal1966} and ATLAS-REFCAT \citep{ATLASREFCAT2-2018ApJ}. The near-IR and mid-IR magnitudes were taken from 2MASS \citep{Cutrietal2MASS2003} and ALLWISE \citep{CutriWISE2014}. We used effective temperature priors based on the spectral type classifications reported in SIMBAD \citep{WengerSIMBAD2000}, and the 3D \texttt{mwdust} extinction estimates as extinction priors. The widths of the temperature and extinction prior errors were $\pm1000$ K and $\pm0.1$ mag.  We obtain the age and mass constraints by matching the luminosity and temperature estimates from the SED fits to the \texttt{PARSEC} isochrones to within $1\sigma$.

\begin{figure}
    \centering
    \includegraphics[width=\linewidth]{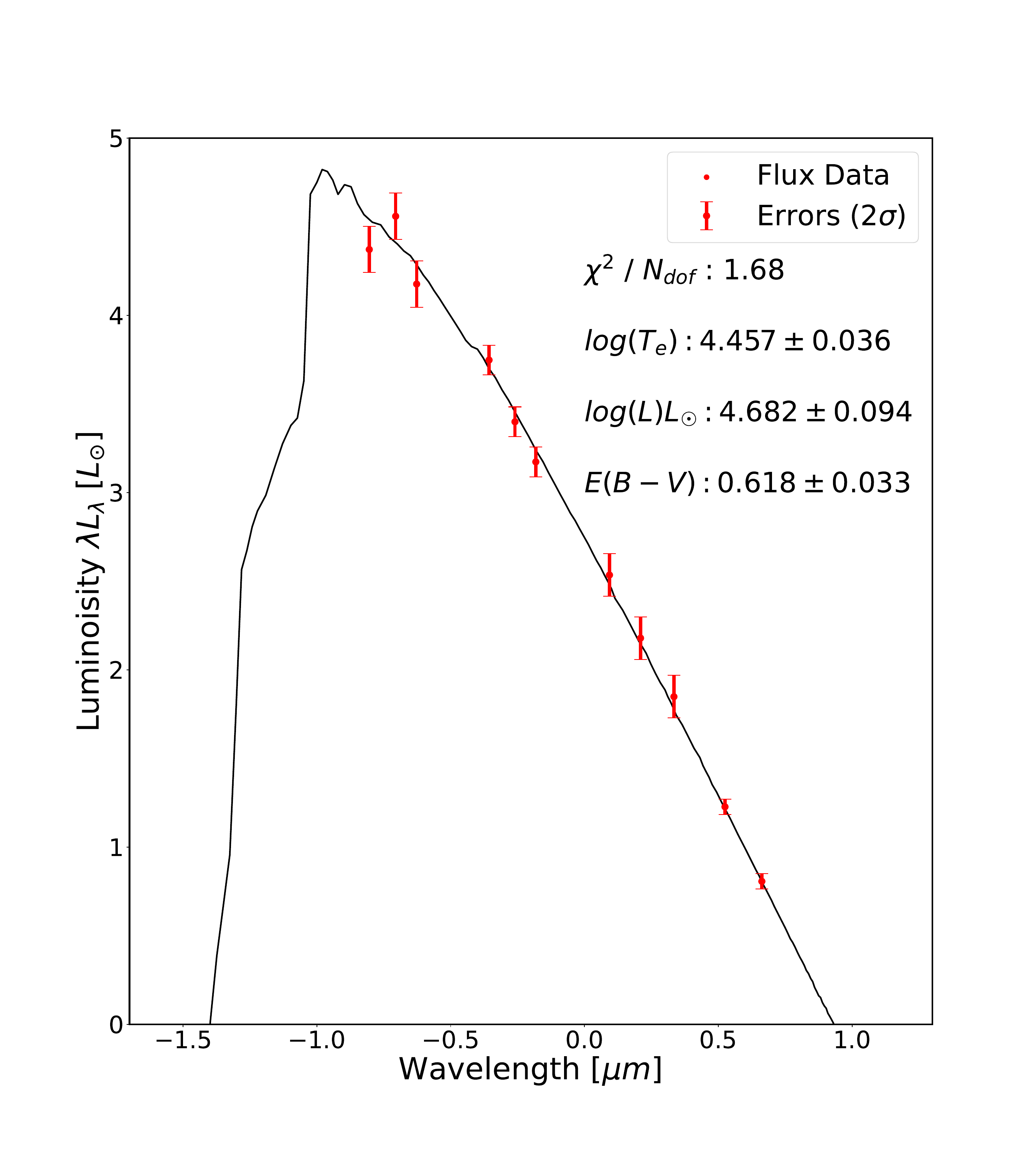}
    \caption{The SED of the eclipsing binary, B8, star ET Tau.}
    \label{fig:SEDfit_ETTAU} 
\end{figure}

\begin{figure}
    \centering
    \includegraphics[width=\linewidth]{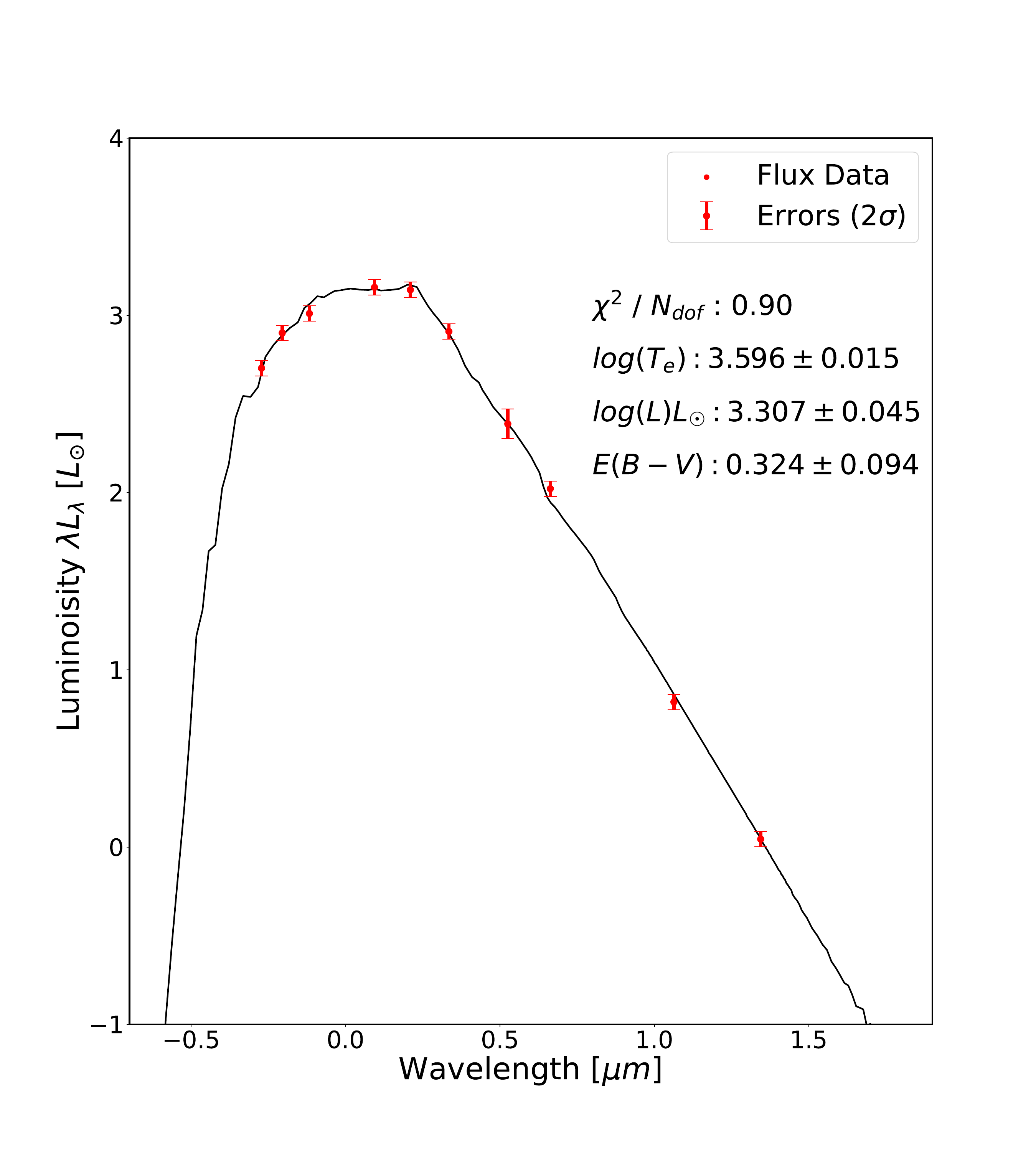}
    \caption{The SED of BD +26935, a spectroscopic binary with a M2/3 primary star.}
    \label{fig:SEDfit_BD26935} 
\end{figure}

\begin{figure}
    \centering
    \includegraphics[width=\linewidth]{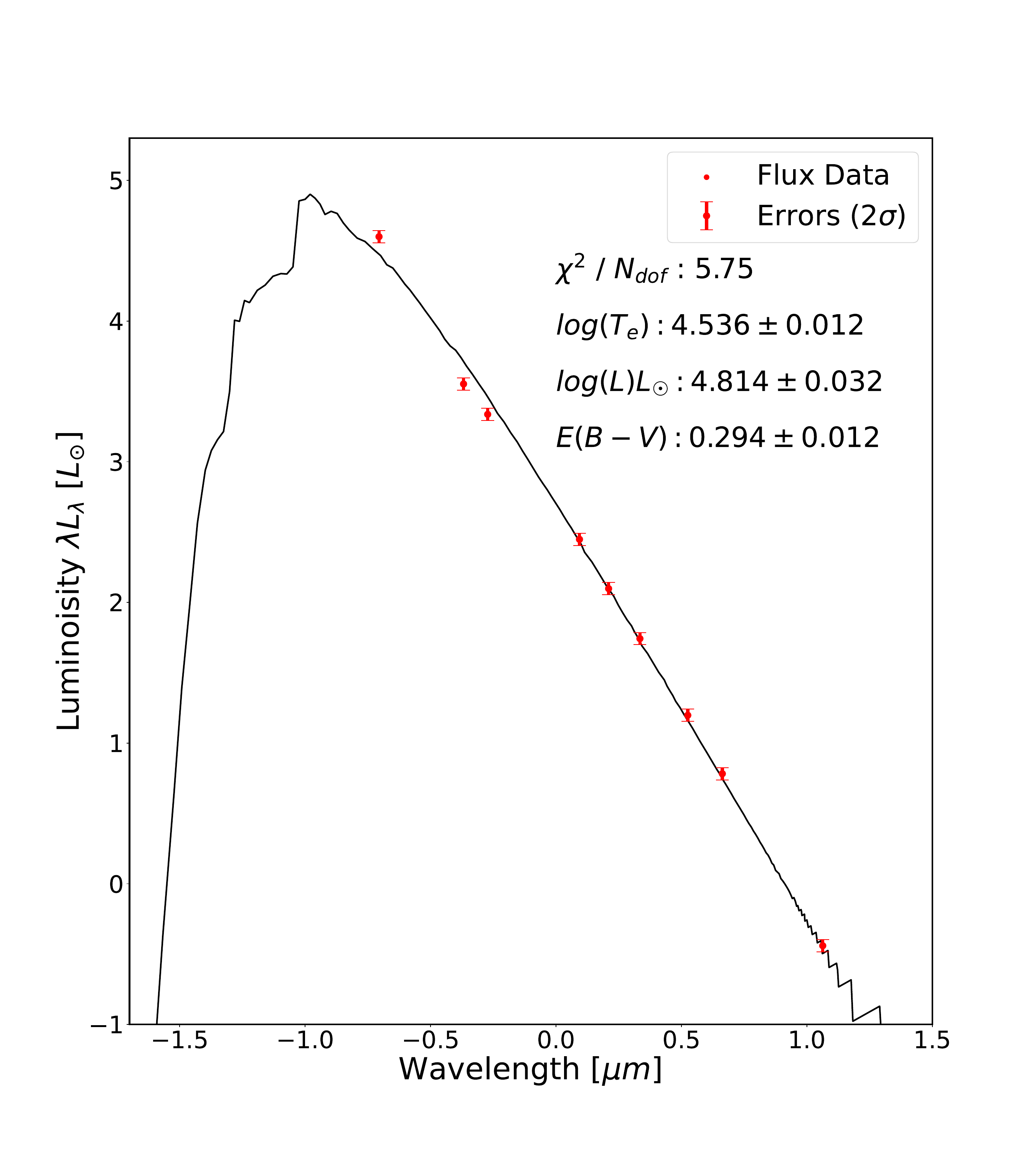}
    \caption{The SED of HD 37366, a spectroscopic binary with an O-type primary star.}
    \label{fig:SEDfit_HD37366} 
\end{figure}

\begin{figure}
    \centering
    \includegraphics[width=\linewidth]{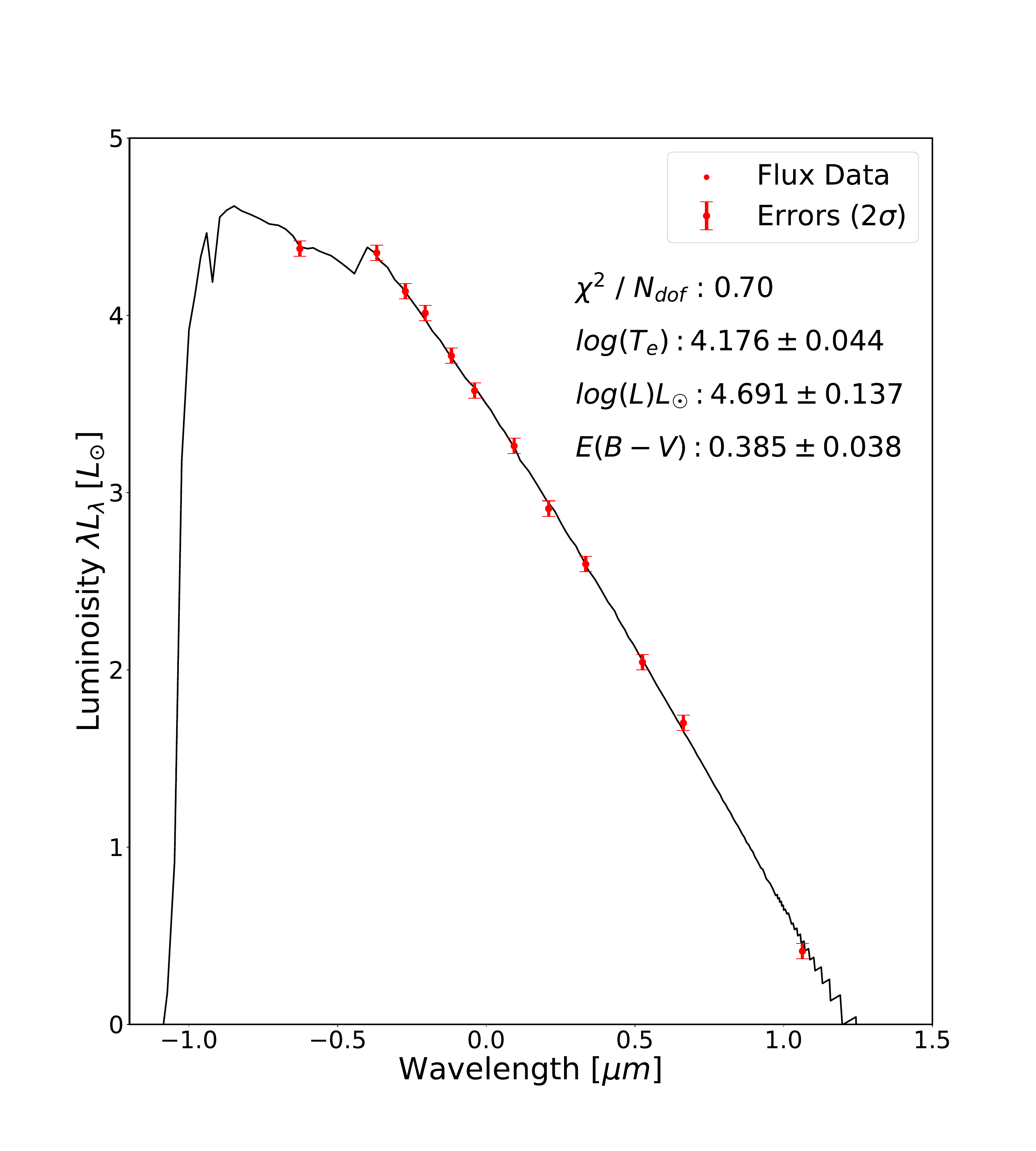}
    \caption{The SED of the B2 IV-V star HD 37367.}
    \label{fig:SEDfit_HD37367} 
\end{figure}

\begin{figure}
    \centering
    \includegraphics[width=\linewidth]{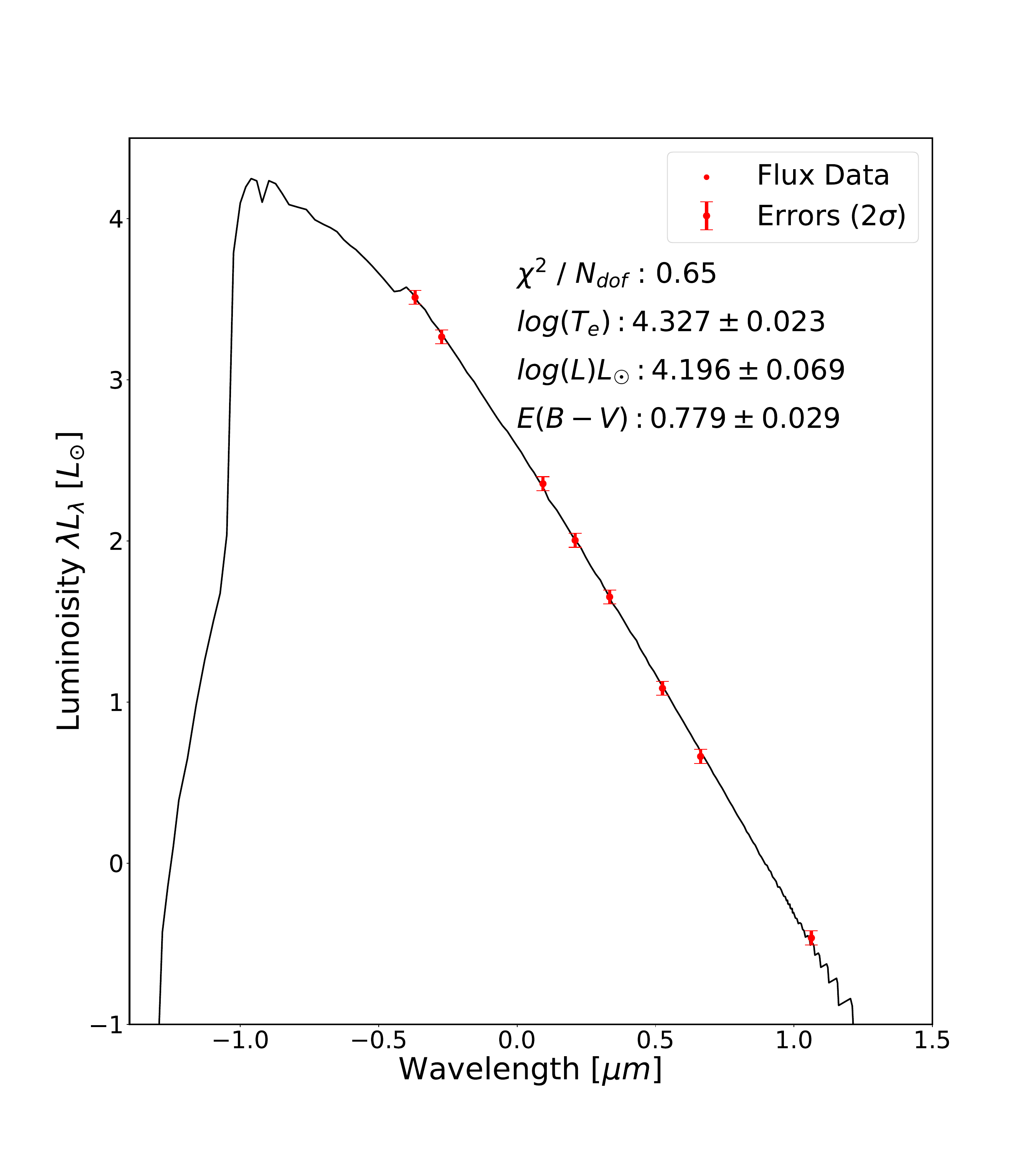}
    \caption{The SED of HD 248666, a B star.}
    \label{fig:SEDfit_HD248666} 
\end{figure}

\begin{figure}
    \centering
    \includegraphics[width=1\linewidth]{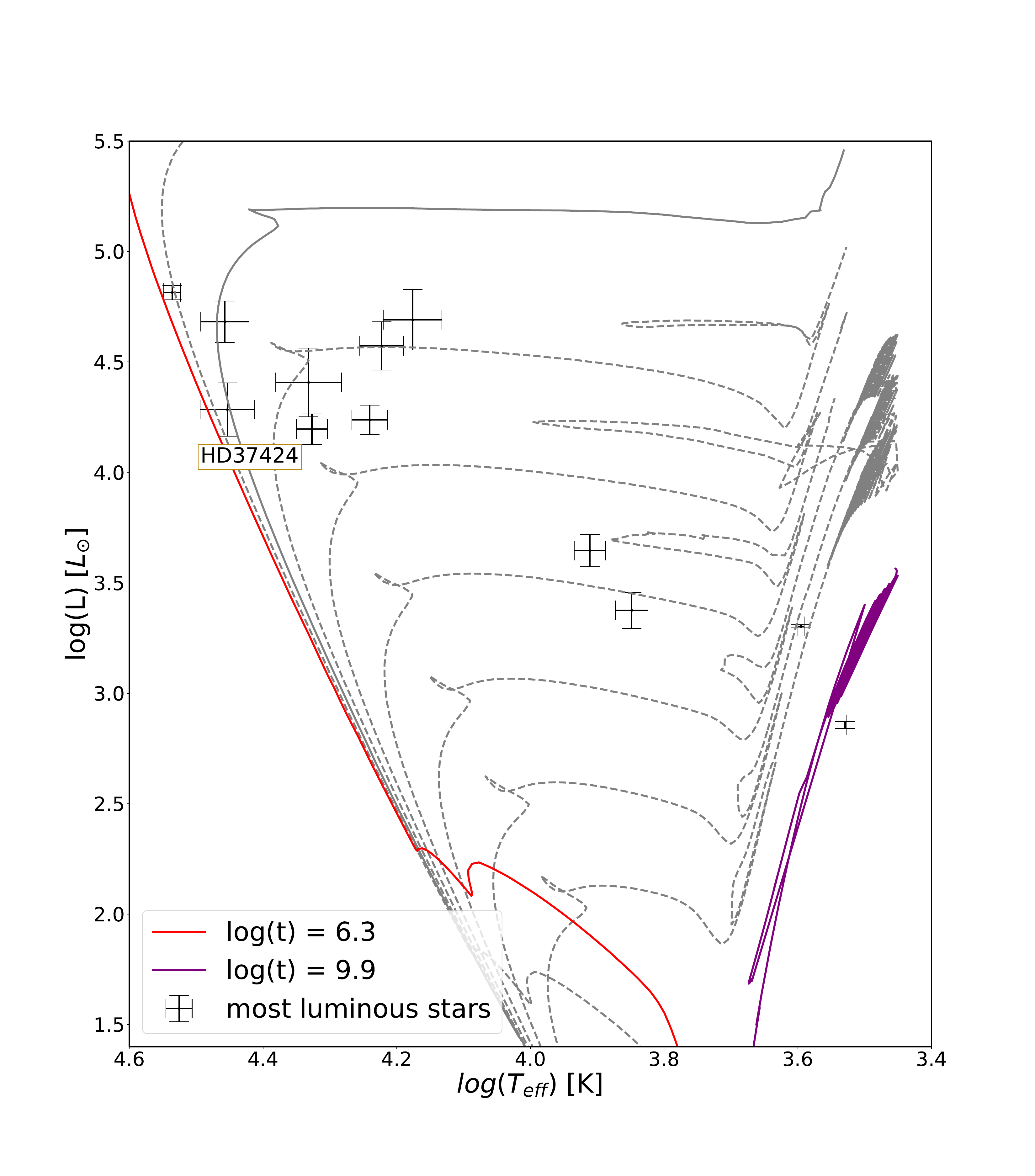}
    \caption{The temperatures and luminosities of the 12 $M_{G}<-3$ mag stars as compared to Solar metallicity \texttt{PARSEC} isochrones shown from $\log_{10}(t/yr) = 6.3$ (solid red, top left) to $\log_{10}(t/yr) = 9.9$ (solid purple, lower right) in steps of $0.3$ dex. See Table \ref{tab:sedlumstars} for numerical results.}
    \label{fig:SED_HR_diagram} 
\end{figure}

Figures \ref{fig:SEDfit_ETTAU}-\ref{fig:SEDfit_HD248666} show the SED fits to the 5 newly modeled systems. With the exception of HD~37366, the fits are generally quite good and for HD~37666 the likely cause is simply underestimated photometric uncertainties. There is no difficulty finding solutions whose temperatures are consistent with the spectral types. ET~Tau is of particular interest because there is a complete eclipsing binary model of the system by \cite{Williamon2016ETTAU}.   They found primary (secondary) masses, luminosities, and temperatures of $(14.34\pm0.28)\msun$ ($6.34\msun\pm0.12\msun$), $10^{4.5\pm0.02}\lsun$ ($10^{3.8\pm0.02}\lsun$), and $10^{4.48\pm0.002}$~K ($10^{4.18}$K). This agrees reasonably well with our estimates fitting only a 
single star to the SED. We modestly overestimate the luminosity ($10^{4.68\pm0.09}L_\odot$), obtain a consistent temperature ($10^{4.46\pm0.04}$~K), and a consistent primary mass $15.12\msun-18.19\msun$.  

Table \ref{tab:sedlumstars} has the estimated ages, masses, luminosities, temperatures, known or unknown spectral classifications, transverse distances from S147, and the $\chi^2$ per degree of freedom of the fits. Figure \ref{fig:SED_HR_diagram} shows the temperatures and luminosities of these stars on a Hertzsprung-Russell diagram. There are 8 stars more luminous than $10^4 L_\odot$, all of which are hot O and B stars. The most luminous is the O star HD~37366 which has an estimated mass of $20.0-21.7M_\odot$. It is a spectroscopic binary, so the luminosity is likely modestly overestimated. HD~37367, is the second most luminous, with an estimated mass of $13.5-15.5M_\odot$, followed closely by the eclipsing binary ET~Tau. The unbound binary companion, HD~37424, is the sixth most luminous, with an estimated average mass of $\sim13.52\msun$. The rest have modestly lower luminosities and masses. The red giants all have significantly lower luminosities because the bolometric correction going from the G band to the total luminosity is much smaller than for the hot stars with SEDs peaking in the ultraviolet.

\section{Progenitor Mass Analysis} \label{sec:analysis}

The next step is to constrain the age distribution of the stars. We construct two sets of 13 stellar density maps in magnitude and color, $F_{jk}(t_i) = F_{jk}^{i}$, similar to Hess diagrams. Here, the index $i$ denotes the age bin, while $j$ and $k$ index the $G$-band absolute magnitude and the $B_P - R_P$ color, respectively. One set of maps consists of single stars and the other of non-interacting binary stars. For both sets, the 13 age bins ($t_{i},$ where $i= 1...13$) range from $10^{6.3}-10^{6.6}$ years to $10^{9.8}-10^{10.1}$ years, sampled in 0.3 dex intervals using the \texttt{PARSEC} isochrones (see Table \ref{tab:singlemcmcresults}). We then build an observed stellar density map by placing each selected star in its corresponding [$j$, $k$] bin based on its extinction-corrected Gaia photometry. To estimate the best-fit age distribution, we fit the observed density map with each of the two model sets using a Monte Carlo Markov Chain (MCMC) likelihood approach with the \texttt{emcee} package \citep{Foreman-Mackey2013emceeMCMC}

To compute the likelihood, we start with the Poisson probability for the number of stars found in each [$j$, $k$] bin. Let $N^{\star}_{jk}$ be the number of observed stars in each cell with $\sum_{jk} N^{\star}_{jk} = N^{\star} = 439$. The number of modeled stars in each bin is $N_{jk} = \sum_i \alpha_i F^{i}_{jk}$, where $\alpha_{i}$ is proportional to the star formation rate of each age bin $i$. The total number of modeled stars is $N = \sum_{jk} N_{jk}$. To avoid undefined or non-physical values in bins with zero predicted stars (i.e., $\ln(0)$), we assign a small value of $1\times10^{-32}$ to empty cells. The Poisson probability for each bin is

\begin{equation} \label{equ:poissonprobequ}
    \frac{N_{jk}^{N_{jk}^{\star}} e^{-N_{jk}}}{N_{jk}^{\star}!}
\end{equation}

\noindent and the logarithm of the likelihood for all $N^{\star}$ stars is

\begin{equation} \label{equ:loglikelihood}
    \ln L = \sum_{jk}^{with} \ln \bigg({r N_{jk}^{N_{jk}^{\star}}} \bigg) - \sum_{jk}^{all} r N_{jk} ,
\end{equation}

\noindent where $r$ is a re-normalization factor. The first term (``with") is the sum over bins containing stars, the second term (``all") is the sum over all bins, and the factorial $N_{jk}^{\star}!$ is ignored because we only need likelihood ratios. If we optimize the likelihood with $r \equiv 1$, then the likelihood would include Poisson fluctuations in the total model count $N$ relative to $N^{\star}$. Since we only care about how the $N^{\star}$ stars are distributed across age bins, we define

\begin{equation} \label{equ:renormalization}
    r = N^{\star} \bigg[ \sum_{jk} \sum_{i} \alpha_{i} F_{jk}^{i} \bigg]^{-1} ,
\end{equation}

\noindent and apply this renormalization as $\rho_i = r \alpha_i$, such that $N^{\star} \equiv N \equiv \sum_{jk}\sum_{i}\rho_{i}F_{jk}^{i}$. This transforms the likelihood into a multinomial distribution for assigning $N^{\star}$ stars across the age bins.
To prevent numerical divergences (e.g., $\log N_i \to -\infty$) in bins where little to no stars are found, we include a weak prior of

\begin{equation} \label{equ:mcmcpriors}
     - \lambda^{-2}\sum_{i}\bigg[\ln\bigg(\frac{\alpha_{i}\Delta t_{i+1}}{\alpha_{i+1}\Delta t_{i}}\bigg)\bigg]^{2} -\lambda^{-1}\sum_{i}\bigg[\ln\bigg(\frac{\alpha_{i}}{\alpha_{0}}\bigg)\bigg]^{2}  , 
\end{equation}

\noindent with $\lambda = \ln{10^{3}} = 6.91$. The first term penalizes variations in SFR by a factor of 1000 between adjacent bins, while the second term penalizes deviations from a uniform distribution of the observed stars across the age bins. 

We fit the observed distribution of stars with both the single-star and non-interacting binary models. The single star models draw stellar masses randomly from a Salpeter IMF between $1\msun$ and $100\msun$, and stellar ages uniformly drawn from the Solar metallicity \texttt{PARSEC} isochrones for that age bin, sampled in intervals of $\Delta \log t = 0.01$ dex \citep{PARSEC2012Bressan, PARSEC2013MarigoAGB, PARSEC2020PastorelliTPAGB}. We use $N_{trial} = 3\times 10^{8}$ trial stars per age bin and add a 1 to the cell [$j$, $k$] if the star falls within $0.0 > M_G > -8.0$ and $-1.0 < B_P - R_P < 3.5$. We added a Gaussian extinction uncertainty of $\sigma_{E(B-V)} = 0.01$ mag. The [$j$, $k$] cells have widths of $\Delta M_{G} = 0.04$ mag and $\Delta (B_{p}-R_{p}) = 0.0225$ mag. For the non-interacting binary star models, after finding the mass $M_{1}$ and age of the primary, a secondary mass $M_2$ is randomly drawn from the same isochrone with $1\msun \leq M_{2} \leq M_{1}$ uniformly in mass. We sum the fluxes of the two stars to get the color and magnitudes of the binary. We exclude both model and actual stars outside the color and magnitude ranges of the grid. We found in \cite{cruzcruz2024} that this had little effect on the results.

Both single and binary models are generated assuming a constant star formation rate (SFR) within each age bin with

\begin{equation}
    \frac{dN}{dMdt} = \frac{(x-2)SFR}{M^{2}_{min}}\Bigg(\frac{M}{M_{min}}\bigg)^{-x} ,
\end{equation} \label{equ:SFRequ}

\noindent for a $M > M_{min}$ and a Salpeter IMF with $x = 2.35$. This gives a mean stellar mass $\langle M \rangle = (x-1) M_{min}/(x-2) = 3.86\msun$ for $M_{min} = 1\msun$. Each bin spans a logarithmic time interval $[t_{min,i}, t_{max,i}]$, with width $\Delta t_i = t_{max,i} - t_{min,i}$. Assuming a constant $SFR_i$ within a bin, the number of stars formed above $M_{min}$ is $N_i = SFR_i \Delta t_i / \langle M \rangle$. Therefore, the number of stars that die in a short time interval $\delta t$ today is

\begin{equation}
     N_{i} \frac{\delta t}{\Delta t_{i}} \biggr[ \bigg(\frac{M(t_{min,i})}{M_{min}}\bigg)^{(1-x)} - \bigg(\frac{M(t_{max,i})}{M_{min}}\bigg)^{(1-x)} \biggr] = N_{i}S_{i}\delta t 
\end{equation} \label{equ:starsthatdieindtequ}

\noindent where $M(t)$ is the most massive star still alive at time $t$, and  $S_{i}\delta t$ is the fraction of $M>M_{min}$ stars in bin $i$ that died in the last $\delta t$ years. The derivation of Equation \ref{equ:starsthatdieindtequ} is provided in Appendix B of \cite{cruzcruz2024}.

We calculate the number of deaths in a time range $\delta t$ with $N_{i}S_{i}$ to find the probability that the progenitor of the SNR S147 died from one of the age bins using

\begin{equation} \label{equ:probsintegralequ}
        \frac{P_{i}}{P_{tot}} = \frac{N_{i}S_{i}}{\sum_{all}N_{i}S_{i}} ,
\end{equation}

\noindent which is independent of $\delta t$ and has a total probability of unity.

We optimize the likelihood (Equation \ref{equ:loglikelihood}) and estimate the posterior distributions using the \texttt{emcee} MCMC sampler \citep{Foreman-Mackey2013emceeMCMC}, to determine the values of $\log \alpha_i$. We use 300 walkers, each with a chain length of $10,000$, and discard the first 1,000 samples as the burn-in period. In each MCMC step, we apply the renormalization $\alpha_i \to \rho_i$ (Equation \ref{equ:renormalization}) before calculating the likelihood. The resulting MCMC chain provide the posterior distributions for $\rho_i$ needed for the statistics shown in Figures \ref{fig:number_vs_agebin_combined}–\ref{fig:integral_prob_S147_combined}.

\section{Results} \label{sec:results}

\begin{figure}
    \centering
    \includegraphics[width=\columnwidth]{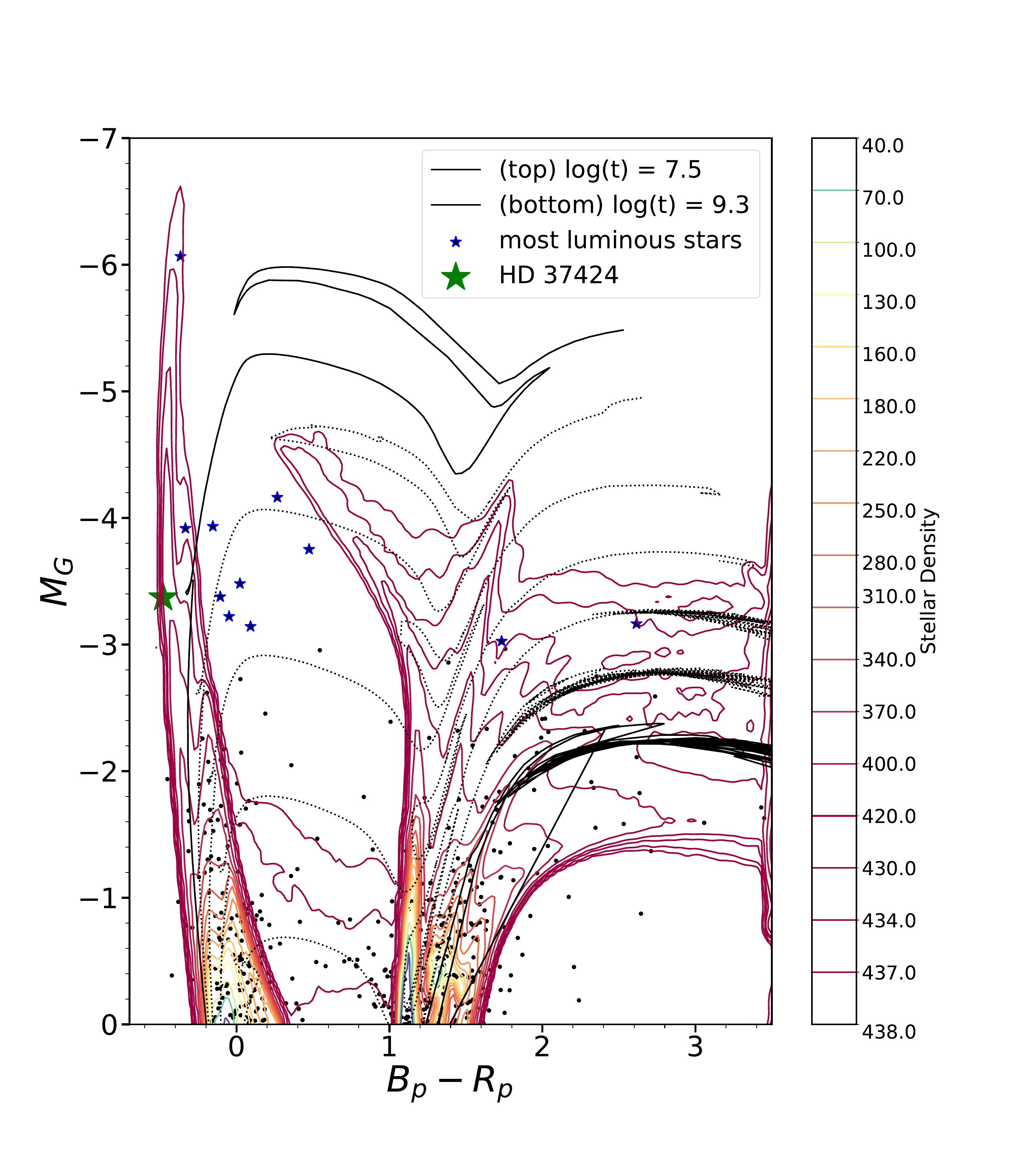}
    \caption{The black dots show the extinction corrected \textit{Gaia} CMD of stars near S147, and the curves are Solar metallicity \texttt{PARSEC} isochrones with ages from $10^{7.5}$(top solid black curve) to $10^{9.3}$ yr (bottom solid black curve) in 0.3 dex steps. The contours are the best-fit single-star density model. The levels are coded by the number of enclosed stars. The large blue stars are the most luminous stars near the S147 SNR, and HD 37424 is the (larger) green star.}
    \label{fig:single_CMD_MCMC_results} 
\end{figure}

\begin{figure}
    \centering
    \includegraphics[width=\columnwidth]{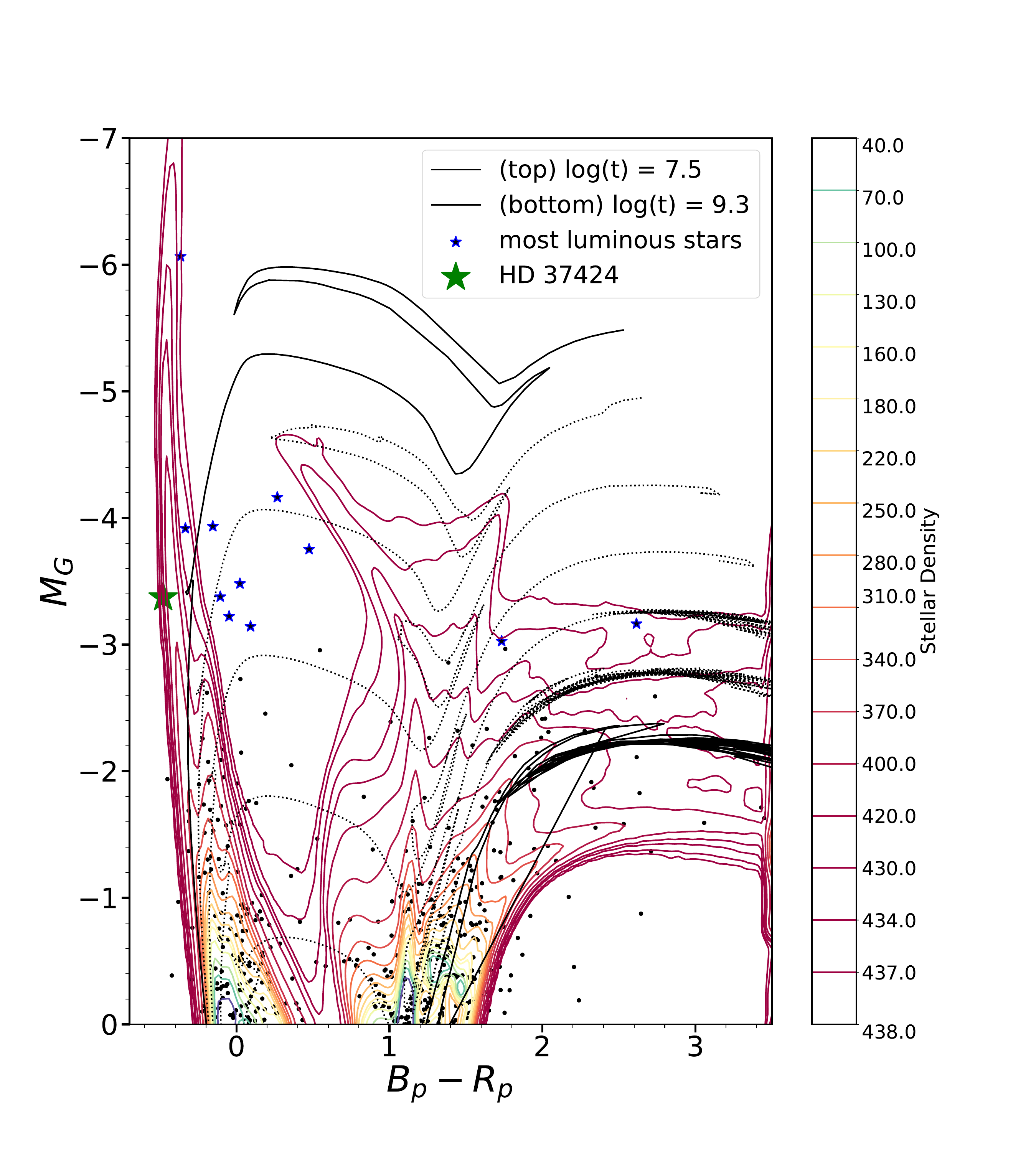}
    \caption{As in Figure \ref{fig:single_CMD_MCMC_results}, but for the best fitting binary model.}
    \label{fig:binary_CMD_MCMC_results} 
\end{figure}

\begin{figure} 
    \centering
    \includegraphics[width=\linewidth]{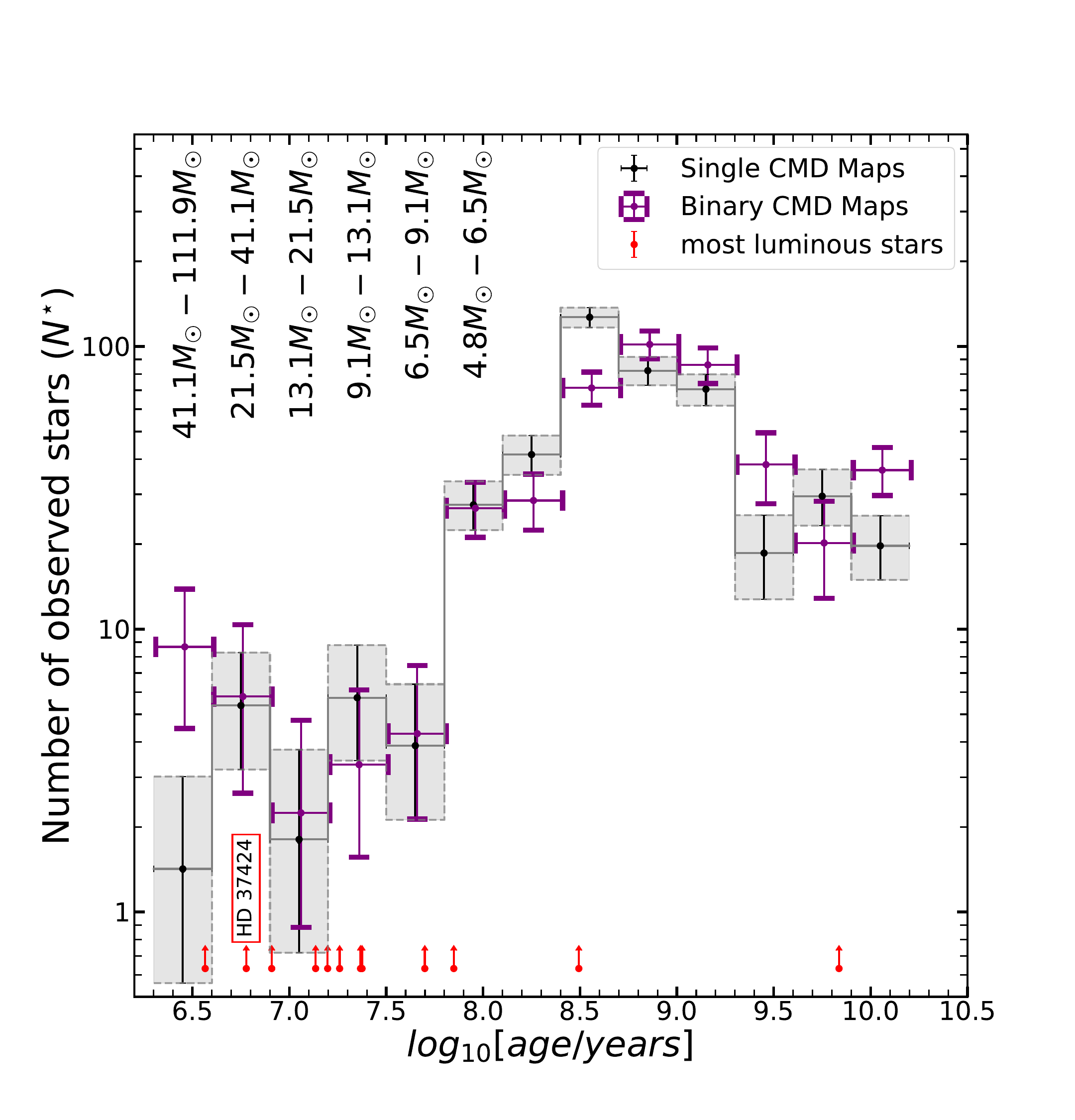}
    \caption{The number of observed stars assigned to each age bin for the single (thin, black) and binary (thick, purple) models. We show the median and the 16 and 84 percentile ranges with points and vertical error bars. The horizontal bins are the widths of the age bins. The (red) arrows are the estimated ages of the individually modeled luminous stars with HD 37424 labeled.}
    \label{fig:number_vs_agebin_combined} 
\end{figure}

\begin{figure}
    \centering
    \includegraphics[width=\linewidth]{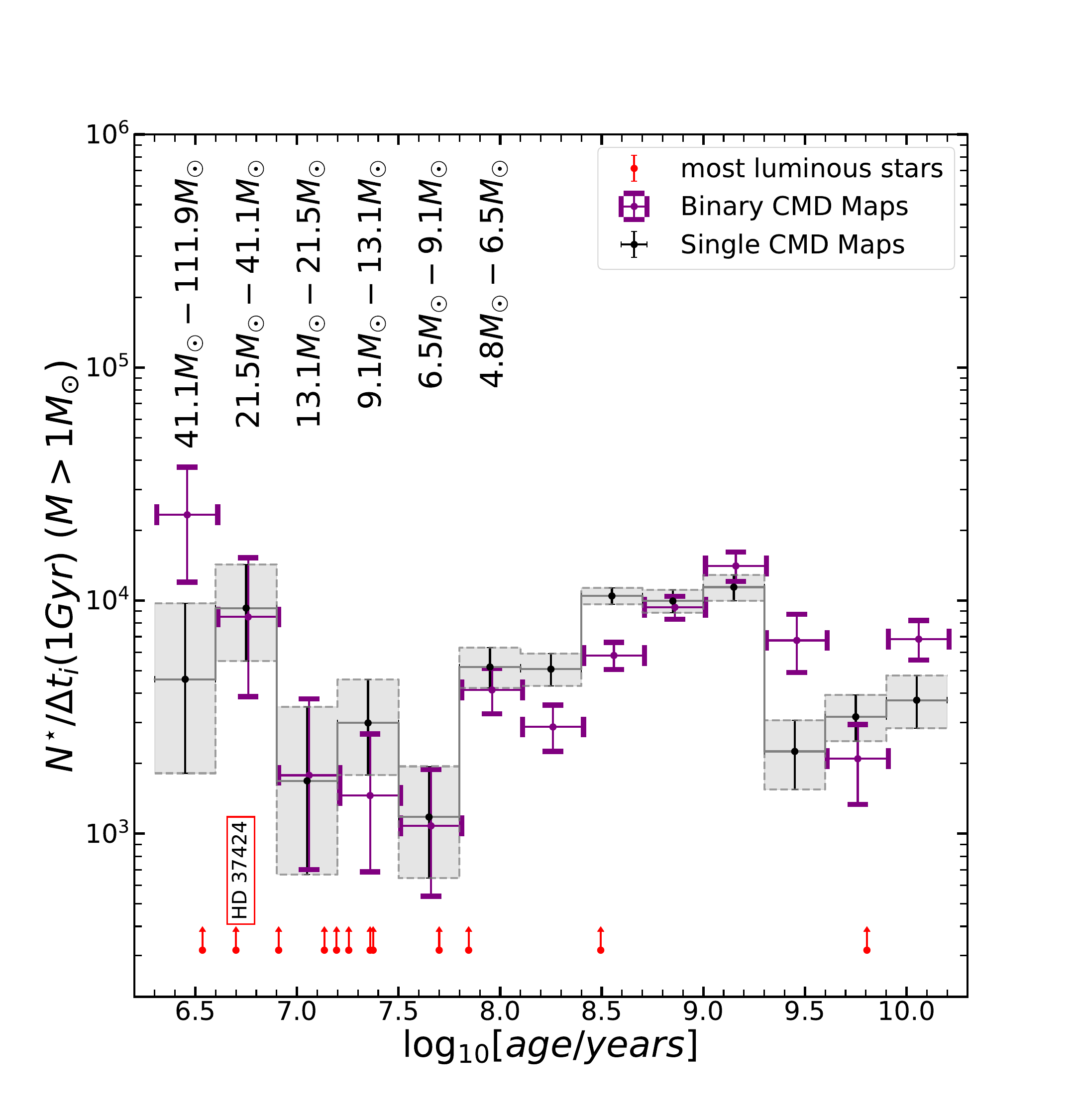}
    \caption{The number $N_{i}^{\star}$ of $M>1\msun$ stars formed in each age bin per $10^{9}$ yr for the single (shaded error bars) and binary (thick purple error bars) models. This is the number of observed stars shown in Fig. \ref{fig:number_vs_agebin_combined} divided by the fraction of the Monte Carlo trials leading to a star on the density grid and normalized to have the SFR constant for a fixed $10^{9}$yrs. The horizontal errors are the 0.3 dex widths of the age bins. The (red) arrows near the bottom are the estimated ages of the most luminous stars with the unbound binary companion labeled HD 37424. }
    \label{fig:Ni_1Gyr_vs_logage} 
\end{figure}

Figures \ref{fig:single_CMD_MCMC_results} and \ref{fig:binary_CMD_MCMC_results} present the density contours of the best-fitting single and binary star models. The single stellar density models fit the extinction-corrected Gaia stellar density map better than the binary star models, but the difference is very sensitive to the floor value we use for the density maps. We present two methods for visualizing the distribution of stars within the stellar density map age bins in Figures \ref{fig:number_vs_agebin_combined} and \ref{fig:Ni_1Gyr_vs_logage}. Figure \ref{fig:number_vs_agebin_combined} displays the model distribution for the $N^{\star} = 439$ observed stars in age (Eqn. \ref{equ:renormalization}). The youngest age bin ($10^{6.3}-10^{6.6}$ yr) includes $\sim 1$ star under the single-star model and $\sim 8$ stars under the binary model. SED fits of the twelve most luminous stars indicate that the unbound binary companion, the eclipsing binary ET Tau, and the spectroscopic binary HD 37366 have ages younger than $< 10^{7.1}$ years and masses of approximately $\sim 13.5\msun$, $\sim 16.7\msun$, and $\sim 20.9\msun$. Binary stellar models find more young luminous stars in the two youngest age bins than single stellar models. Table \ref{tab:singlemcmcresults} and Table \ref{tab:binarymcmcresults} provide the numerical results from the single and binary models. Two of the most luminous stars fall within the third youngest bin ($10^{6.9}-10^{7.2}$ yr), although the single-star model finds almost two, $\sim1.8$, stars in this bin and the binary star model finds $\sim2.2$ stars. 


The distribution of the observed stars is not trivially related to the relative star formation rates of the age bins. In Figure \ref{fig:Ni_1Gyr_vs_logage} we show the number of $M>1\msun$ stars that would be formed in $10^{9}$ yrs given the implied SFR of the bin for both the single and binary stellar density model maps. To calculate these values, we divide the predicted number of observed stars in each age bin (from Fig. \ref{fig:number_vs_agebin_combined}) by the fraction of Monte Carlo trials from the corresponding stellar density model that led to a star on the $F^{i}_{j,k}$ grid and by the temporal width of each bin. Rather than showing a rate, we multiply by $10^{9}$ yrs to show the number of $M>1\msun$ stars that would be formed over this time period. In both models, we see a recent burst of star formation.

\begin{table*}
\centering
    \caption{The MCMC results from using single stellar density model maps. The distribution of the observed stars $N^{\star}$ (column 2, Fig. \ref{fig:number_vs_agebin_combined}) across the age bins (column 1), the implied star formation rate (column 3, Fig. \ref{fig:Ni_1Gyr_vs_logage}), and the probability of stars dying in the last $10^{5}$ yr (column 4, Fig. \ref{fig:Ndt_1e5yrs_vs_agebin_combined}).}
    \setlength{\tabcolsep}{12pt}
        \begin{tabular}{c c c c}
        $t_{min}-t_{max}$ [yrs] & $N^{\star}$ & $N^{\star}/\Delta t_{i}$ ($M>1\msun/1$Gyr) & $N_{i}S_{i}\delta t$ \\ 
        \hline
        $10^{6.3}-10^{6.6}$ & $ 1.42\pm1.23$ & $4595\pm3968$ & $0.0023\pm0.0019$ \\ 
        $10^{6.6}-10^{6.9}$ & $5.37\pm2.55$ & $9283\pm4399$ & $0.0086\pm 0.0040$\\ 
        $10^{6.9}-10^{7.2}$ & $1.81\pm1.52$ & $1684\pm1417$ & $0.0026\pm0.0021$\\
        $10^{7.2}-10^{7.5}$ & $5.73\pm2.68$ & $2983\pm1396$ & $0.0059\pm 0.0027$ \\ 
        $10^{7.5}-10^{7.8}$ & $3.87\pm2.14$ & $1178\pm650$ & $0.0035\pm0.0019$ \\ 
        $10^{7.8}-10^{8.1}$  & $27.54\pm5.46$ & $5187\pm1029$ & $0.0203\pm0.0040$ \\ 
        $10^{8.1}-10^{8.4}$  & $41.47\pm6.65$ & $5079\pm815$ & $0.0258\pm0.0041$ \\
        $10^{8.4}-10^{8.7}$  & $126.83\pm10.18$ & $10465\pm840$ & $0.0694\pm0.0055$ \\ 
        $10^{8.7}-10^{9.0}$  & $82.13\pm9.39$ & $9956\pm1137$ & $ 0.0879\pm0.0100$ \\ 
        $10^{9.0}-10^{9.3}$  & $70.61\pm9.02$  & $11418\pm1458$ & $0.1769\pm0.0225$ \\ 
        $10^{9.3}-10^{9.6}$  & $18.59\pm6.28$ & $2252\pm761$ & $0.0443\pm0.0149$ \\
        $10^{9.6}-10^{9.9}$  & $29.55\pm6.74$ & $3172\pm723$ & $0.0661\pm0.0150$ \\ 
        $10^{9.9}-10^{10.1}$  & $19.72\pm5.13$  & $3737\pm972$  & $0.0588\pm0.0153$ \\
        \hline
        \end{tabular}
    \label{tab:singlemcmcresults}
\end{table*}

\begin{figure}
    \centering
    \includegraphics[width=\linewidth]{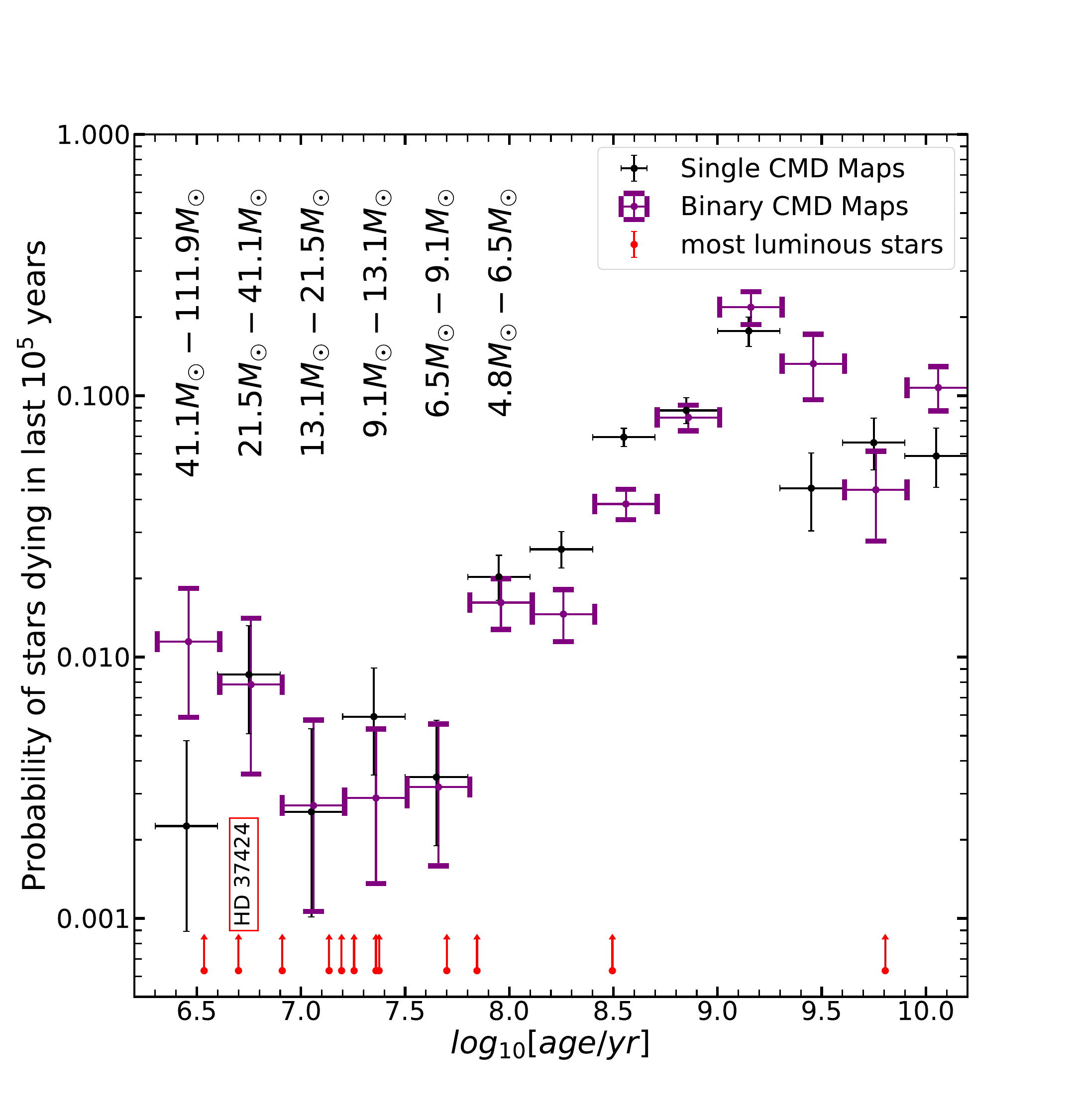}
    \caption{The probability of the number of stellar deaths over the last $\delta t_{i} = 10^{5}$ yr for the single (thin, black) and binary (thick, purple) models. The horizontal error bars span the 0.3 dex age bin widths. The red arrows show the ages of the most luminous stars including the unbound binary companion, HD 37424.}
    \label{fig:Ndt_1e5yrs_vs_agebin_combined} 
\end{figure}

\begin{figure}
    \centering
    \includegraphics[width=\linewidth]{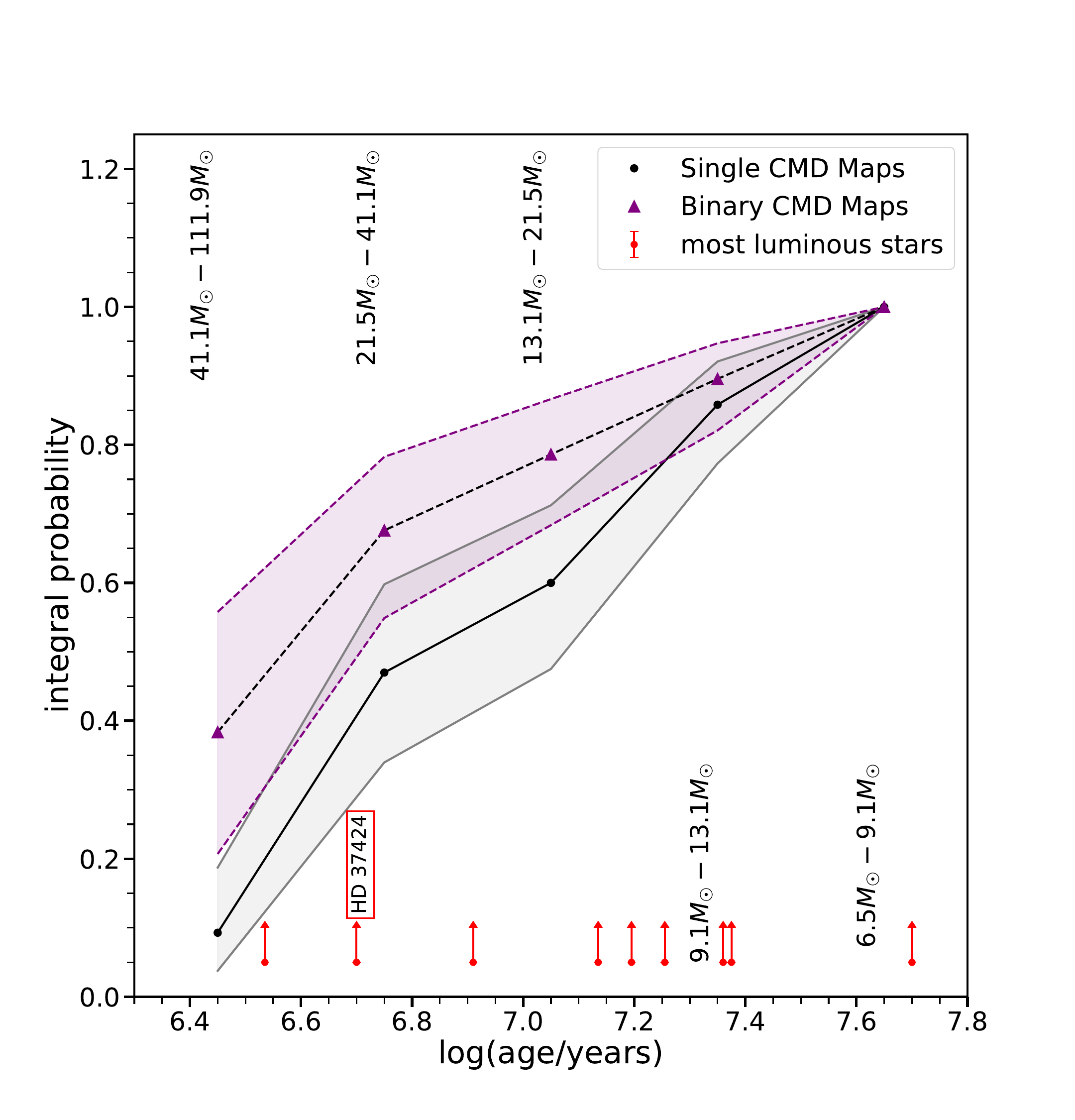}
    \caption{The integral probability distribution of the number of stellar deaths as a function of age. The solid, grey (dashed, purple) lines are the $1\sigma$ confidence range for the median number of stellar deaths within each age bin found for the single (binary) model. The age estimates of the most luminous stars are shown as arrows, near the bottom, with S147 SNR's unbound binary companion labeled, HD 37424. Both distributions are truncated at the oldest age bin that can produce a ccSN, excluding the explosion of a merger remnant.}
    \label{fig:integral_prob_S147_combined} 
\end{figure}

Figures \ref{fig:Ndt_1e5yrs_vs_agebin_combined} and \ref{fig:integral_prob_S147_combined} present the differential and integral distributions in age of stars expected to have exploded within the past $\delta t = 10^{5}$ years. The resulting probabilities are low because, in a randomly selected volume with this stellar age distribution, the likelihood of having a supernova is small. However, we are only interested in relative and not absolute probabilities. Our results favor a higher-mass progenitor. The differential probability (Fig. 
\ref{fig:Ndt_1e5yrs_vs_agebin_combined}), resulting from using the single stellar density maps, shows the age bin, $10^{6.6}-10^{6.9}$ yr, corresponding to high-mass stars with mass range $21.5\msun - 41.1\msun$, has a median likelihood roughly 4 times those of higher masses $\geq 41.1\msun$ and 4 times higher than those of lower masses $\leq 21.5\msun$ bins (see Table \ref{tab:singlemcmcresults}). When considering the binary model, we find that the same age bin has a median likelihood roughly $\sim1.4$ times less than that of higher masses $\geq 41.1\msun$, but a median likelihood $\sim 3$ times higher than those with $\leq 21.5\msun$ bins. In the integral probability distributions (Fig. \ref{fig:integral_prob_S147_combined}), the age range $10^{6.6}$yr$-10^{6.9}$yr encompasses $\sim83\%$ ($\sim64\%$) of the probability of the single (binary) star model. The third youngest age bin, $10^{6.9-7.2}$ yr ($13.1\msun-21.5\msun$), we find an integral probability of $\sim52\%$ (binary: $\sim43\%$). 

We know HD 37424's mass is $\sim13.5\msun$, and that it was in a binary with the progenitor of S147. In order for the progenitor to have gone supernovae before HD 37424, it should have been as massive or more massive than $\sim13.5\msun$. Therefore, we use HD 37424's mass as the minimum mass for the progenitor and truncate the probability distribution of the progenitor's age and mass at the age bins that contain stars with masses $>13.1\msun$. Figure \ref{fig:Ndt_1e5yrs_vs_agebin_combined} shows that, in the case of the single stellar models, there is a 4 times higher differential probability that the progenitor came from the mass bin $21.5\msun - 41.1\msun$ than from $13.1\msun-21.5\msun$. We see the same trend after calculating the probabilities using either single or binary models. Both differential and integral probabilities strongly support a high-mass progenitor between $21.5\msun-41.1\msun$. The progenitor mass range is consistent with a more massive star in a binary with HD 37424 before undergoing mass loss/mass transfer, exploding, disrupting the system, and leaving an unbound companion.

\begin{table*}
\centering
    \caption{The MCMC results from using binary stellar density model maps. The distribution of the observed stars $N^{\star}$ (column 2, Fig. \ref{fig:number_vs_agebin_combined}) across the age bins (column 1), the implied star formation rate (column 3, Fig. \ref{fig:Ni_1Gyr_vs_logage}), and the probability of stars dying in the last $10^{5}$ yr (column 4, Fig. \ref{fig:Ndt_1e5yrs_vs_agebin_combined}).}
    \setlength{\tabcolsep}{12pt}
        \begin{tabular}{c c c c}
        $t_{min}-t_{max}$ [yrs] & $N^{\star}$ & $N^{\star}/\Delta t_{i}$ ($M>1\msun/1$Gyr) & $N_{i}S_{i}\delta t$ \\ 
        \hline
        $10^{6.3}-10^{6.6}$ & $8.66\pm4.70$ & $23327\pm12666$ & $0.0115\pm0.0062$ \\ 
        $10^{6.6}-10^{6.9}$ & $5.78\pm3.87$ & $8505\pm5701$ & $ 0.0078\pm0.0053$\\ 
        $10^{6.9}-10^{7.2}$ & $2.24\pm1.94$ & $1779\pm1541$ & $0.0027\pm0.0023$ \\
        $10^{7.2}-10^{7.5}$ & $3.32\pm2.27$ & $1456\pm994$ & $0.0028\pm0.0019$ \\ 
        
        $10^{7.5}-10^{7.8}$ & $4.27\pm2.65$ & $1079\pm670$ & $0.0032\pm 0.0019$ \\ 
        $10^{7.8}-10^{8.1}$  & $26.8\pm5.97$ & $4134\pm922$ & $0.0161\pm0.0036$ \\ 
        $10^{8.1}-10^{8.4}$  & $28.5\pm6.49$ & $2868\pm652$ & $0.0146\pm 0.0033$ \\
        $10^{8.4}-10^{8.7}$  & $71.3\pm9.57$ & $5812\pm780$ & $0.0385\pm0.0052$ \\ 
        
        $10^{8.7}-10^{9.0}$  & $101.6\pm11.4$ & $9353\pm1052$ & $0.0826\pm0.0093$ \\ 
        $10^{9.0}-10^{9.3}$  & $86.1\pm12.4$  & $14076\pm2027$ & $0.2181\pm0.0314$ \\ 
        $10^{9.3}-10^{9.6}$  & $38.2\pm10.8$ & $6746\pm1915$ & $0.1326\pm0.0376$ \\
        $10^{9.6}-10^{9.9}$  & $20.2\pm 7.74$ & $2095\pm803$ & $0.0437\pm0.0167$ \\ 
        $10^{9.9}-10^{10.1}$  & $36.5\pm7.10$  & $6829\pm1327$  & $0.1074\pm0.0209$ \\
        \hline
        \end{tabular}
    \label{tab:binarymcmcresults}
   
\end{table*} 

\section{Discussion} \label{sec:discussion}


We examine the properties of the 439 stars with $-8 < M_{G}<0$ mag and $-1.0< B_{p}-R_{p}< 3.5$ in a volume surrounding the S147 SNR. These stars lie in a cylinder with a projected radius of $100$pc from HD 37424 and a length along the line of sight of $\sim358$ pc. We analyze the SEDs of the twelve most luminous stars in detail. We find that the two most luminous stars, HD 37424 and HD 37367 have luminosities, masses, and ages of roughly $10^{4.3\pm0.1}\lsun$ ($10^{4.69\pm0.14}\lsun$), $13.5\msun\pm0.1\msun$ ($14.30\msun \pm 0.09\msun$), and $10^{6.8\pm0.01}$yr ($10^{7.10}$yr$-10^{7.17}$yr).


We created two sets of stellar density models with single and binary stars. We modeled the age distribution of the observed stars and the age distribution of stars that have recently (within the past $10^{5}$yrs) died for both models. We find evidence that the progenitor of S147 was a young high-mass star ($21.5\msun - 41.1\msun$) with an age ($10^{6.6}-10^{6.9}$ yr) that is consistent with the age of HD 37424 ($10^{6.3}-10^{7.10}$ yr), the pre-explosion binary companion that became unbound by the supernovae event that produced S147 \citep{Dincel2015DiscoveryOBRunawayS147}. We find similar results when using different single ($\sim83\%$) and non-interacting binary ($\sim64\%$) stellar population models. Our estimates for the range of progenitor mass are consistent with the expected primary mass and age of a progenitor in the unbound binary scenario of S147.


After considering Vela, the Crab, and now S147, we conclude that this method is better suited to statistically study ensembles of SNRs. For a single SNR, it can provide unambiguous results only if the surrounding stellar population is dominated by a single, well-defined starburst. Even with precise spatial and geometric information, the region may still suffer from projection effects, incompleteness, and contamination \citep{Kochanek2022} that can create or conceal such bursts. Stars within the projected region may lie far from the progenitor's birth site, or the region may lack a sufficient number of stars to robustly identify the relevant starburst. 

The method would also benefit from including a ``comparison" sample of stars where there is no SNR. This has been done by \cite{MaozCarles2010} in the LMC and SMC. They analyzed the full stellar catalogs from \cite{HarrisZaritsky2004,HarrisZaritsky2009} and all the SNRs simultaneously. This allows a clean calculation of the probability that a given age bin produces a SN. For ccSNe, the study was limited by incomplete knowledge of the type of supernova explosion, because longer-lived Type I progenitors contribute a real signal at older ages. Comparing the SFHs and age distributions of stellar populations not near a SNR to those near an SNR for populations of remnants in our Galaxy may be a more promising avenue to estimate progenitor masses statistically.   

Applying this approach to a sample of Galactic SNRs is challenging because distances to SNRs are generally more uncertain. For S147, \cite{Kochanek2024DistanceS147} demostrated a new approach to measuring distances by searching for the appearance of high-velocity absorption features in background stars superposed on the SNR as a function of distance. They demonstrate that this method can provide accurate distances to SNRs where there is no parallax for a neutron star remnant or binary companion. With more well-constrained distances to SNRs, it will be possible to carry out such a statistical survey of Galactic SNRs. Some Galactic SNRs that are of particular interest are those which are currently interacting binaries \citep[SS 433, HESS J0632+057, 1FGL J1018.6-5856;][respectively]{SS433-distance-2004, Hinton-Hess2009ApJ, Corbetbinary2011, 1FGLJ10186-dists-Ackermann2012}.


\section*{Acknowledgements}  \label{sec:acknowledgements}
Elvira Cruz-Cruz is supported by NASA FINESST Fellowship 80NSSC23K1444. Christopher S. Kochanek is supported by NSF grants AST-2307385 and AST-2407206. This research has made use of the VizieR catalogue access tool, CDS, Strasbourg, France \citep{10.26093/cds/vizier}. This work has made use of data from the European Space Agency (ESA) mission {\it Gaia} (\url{https://www.cosmos.esa.int/gaia}), processed by the {\it Gaia} Data Processing and Analysis Consortium (DPAC, \url{https://www.cosmos.esa.int/web/gaia/dpac/consortium}). Funding for the DPAC has been provided by national institutions, in particular the institutions participating in the {\it Gaia} Multilateral Agreement. 

ECC thanks her parents, Esteban Cruz-Ledezma and Elvira Cruz-Lopez, who immigrated from Mexico to the US, and taught her how to work the land and reach for the stars. \textit{Gracias Mamá y Papá, los amo mucho.} ECC also thanks Dominick M. Rowan for helpful discussions.

\bibliography{main}{}
\bibliographystyle{aasjournal}

\end{document}